# CHARACTERIZING TRANSFER GRAPHS OF SUSPICIOUS ERC-20 TOKENS

Calvin Josenhans [1], Andrey Kuehlkamp [2], and Jarek Nabrzyski [2]

[1] Department of Computer Science, Indiana University, Indiana, USA
[2] Center for Research Computing, University of Notre Dame, Indiana, USA

## ABSTRACT

*Ethereum is currently the second largest blockchain by market capitalization and a popular platform for cryptocurrencies. As it has grown, the high value present and the anonymity afforded by the technology have led Ethereum to become a hotbed for various cybercrimes. This paper seeks to understand how these fraudulent schemes may be characterized and develop methods for detecting them. One key feature introduced by Ethereum is the ability to use programmable smart contracts to execute code on the blockchain. A common use of smart contracts is implementing fungible tokens with the ERC-20 interface. Such tokens can be used to impersonate legitimate tokens and defraud users. By parsing the event logs emitted by these ERC-20 contracts over 20 different periods of 100K blocks, we construct token transfer graphs for each of the available ERC-20 tokens on the blockchain. By analyzing these graphs, we find a set of characteristics by which suspicious contracts are distinguished from legitimate ones. These observations result in a simple model that can identify scam contracts with an average of 88.7% accuracy. This suggests that the mechanism by which fraudulent schemes function strongly correlates with their transfer graphs and that these graphs may be used to improve scam-detection mechanisms, contributing to making Ethereum safer.*

## 1. INTRODUCTION

A blockchain is a decentralized, mostly pseudonymous, ledger that uses cryptography to ensure security. The most common use case for this technology is in cryptocurrencies, where each transaction is cryptographically signed and stored on the blockchain. Since Bitcoin was introduced in 2008 [1], cryptocurrency has become an area of interest for investors and researchers, and many other blockchains have developed seeking to improve upon Bitcoin's original design.

Ethereum, which launched in 2015, was the first blockchain to offer the ability to write smart contracts in a Turing complete programming language and publish them on the blockchain [2]. This has resulted in the development of a variety of decentralized applications on the Ethereum network that are implemented in smart contract code. Smart contracts are also commonly used to manage tokens. Tokens have a variety of uses, including representing assets, corresponding to membership in an organization, and use as currency [3]. To make it easier to develop and exchange tokens, the ERC-20 standard was adopted by the Ethereum community for fungible tokens; ERC-721 later allowed for non-fungible tokens, and ERC-1155 offers the ability to implement both within one contract.

Due to the pseudonymous nature of blockchains and the significant amount of money present in the ecosystem, these technologies have become targets for scammers seeking to exploit vulnerabilities. Over $85 million worth of ETH and ERC-20 tokens were lost in the first two months of 2024 due to phishing attacks, and scammers continue to target the platform through a variety of means. Researchers have taken various approaches to combat the proliferation of scams through various identification methods and classification models [4], [5], [6], [7]. We investigate the applicability of ERC-20 transfer graphs to scam identification, which to our knowledge is an approach that has not been taken in the literature.

In summary, this paper makes the following contributions:

1. We present a novel approach applying token transfer networks to the detection of fraudulent contracts.
2. We show that the structure of token transfer networks can be correlated to malicious behavior.
3. We present a new dataset with Ethereum ERC-20 token transfers, annotated with labels collected from open-source intelligence (OSINT).
4. We show that characteristics of fraudulent token networks are consistent across numerous time periods.

Both the dataset we created and the code we used to conduct our experiments will be made openly accessible upon publication. The remainder of this paper is organized as follows: an overview of existing research and methods into both Ethereum networks and Ethereum scam detection is presented in section 2. Section 3 describes how we collect over 314 M transfers from 269,779 unique contracts in 20 periods of 100K blocks, and build 632K token transfer networks, extracting summary features from each of these. In section 4 we make inferences from the distributions of these features and evaluate the ability of a logistic regression model to predict the malicious behavior of tokens. Finally, in section 5 we draw conclusions from our findings, and discuss directions for future work.

## 2. BACKGROUND AND RELATED WORK

Ethereum was introduced in 2014 as an alternative to existing blockchains. In addition to facilitating cryptocurrency transactions, it supports the ability for users to write Turing complete code that can be published and executed by anyone on the blockchain [2]. Code written in a high-level programming language is compiled to a bytecode that runs on the Ethereum Virtual Machine (EVM) [8]. This has led Ethereum to become a hotbed for the development of decentralized applications and gain worldwide reach. It is currently the second largest blockchain by market capitalization, behind only Bitcoin, and remains one of the most active blockchains to this day. Unfortunately, *where there is money, there are those who follow it*, and as such scammers and malicious actors have targeted the Ethereum network.

Research into scam detection uses a variety of means to understand and detect these scams. Many approaches seek to identify anomalous account behavior. Aziz et al.[4] and [6] take into account features measuring the frequency, lifetime, and amount of transactions associated with flagged addresses. Farrugia et al.[5] and [7] do the same, and additionally incorporate ERC-20, ERC-721, and ERC-1155 transfers. All 4 approaches achieve strong performance across a variety of machine learning models.

For detecting malicious behavior of smart contracts, a common approach is to augment other activity-based features with features related to the contract's bytecode. Pan et al.[7] and [9] incorporate opcode frequency measures into their models. This approach is also seen in [10], which focuses on contracts that implement Ponzi schemes. They use account features specifically chosen to capture Ponzi behavior in addition to code features corresponding to the frequency of various EVM bytecodes in the contracts to develop a classifier that identifies Ponzi contracts with high accuracy.

Another vehicle for malicious behavior is the creation of scam ERC-20 tokens. On the Uniswap decentralized exchange, researchers found that 50% of listed tokens are linked to known scams by analyzing their transaction patterns [11]. Also prevalent are counterfeit tokens, which mimic more popular tokens and are utilized to defraud users in a variety of scams [12].

A useful perspective for surveying Ethereum that has been applied to scam detection is the interpretation of various aspects of Ethereum activity as a graph [13], [14]. Different graphs constructed by various studies include the graph of transactions, graphs of contract creation and interaction, and the graphs of ERC-20 and ERC-721 transfers.

Research has used transaction data to build graphs of money flow, contract creation, contract interaction, and more. Studies found heavy tailed degree distributions in the graphs of money flow and contract creation in the Ethereum network [14], [15], and approaches have used sliding window techniques to observe growth and evolution of the network and correlate changes in graph properties with external events [15], [16].

Chen et al.[17] builds graphs of token holders and creators, finding power law distributions for both token activity and for degree of the token holder graph. They observe that while 60.3% of creators published only one token contract, some accounts have created over 1000, an extreme imbalance in this particular activity.

Another perspective looks at the graph formed by the transfers of ERC-20 and ERC-721 tokens. Somin et al.[18] finds power law distributions for the popularity of tokens and for the degrees of the ERC-20 graph, and Casale-Brunet et al.[19] finds similar patterns in the transfers of ERC-721 non-fungible tokens, leading to the conclusion that token activity shares many similarities with social networks.

With regards to the transfer graphs of individual tokens, [20] finds that power law degree distributions are less applicable in the cases of individual tokens, noting that these networks are usually dominated by so-called "emitters" and "exchanges", addresses with high out-degree and in-degree respectively. Loporchio et al.[21] analyzes the network topology of the top 100 tokens and observes a low clustering coefficient for all networks, suggesting that these token networks are not small-world networks. They also use clustering algorithms to determine whether the network structure reveals information about the token's purpose but conclude that the network topology does not seem to reflect the contract's role in the broader Ethereum network.

In applying Ethereum graphs to scam detection, [22] uses a network embedding model to extract features and an SVM to identify phishing versus non-phishing nodes. They utilize the local subnetwork consisting of a node's first-order neighbors and their transaction records to develop the embedding. Wan et al.[23] uses features of similar subnetworks in addition to time-series based data to identify phishing accounts, and [24] proposes the use of a different embedding method incorporating time and amount information for the transactions. Chen et al.[25] takes advantage of a graph cascading approach, where the network features of a node's neighbors are used in addition to those of the node itself. The study focuses on the network formed by a subset of accounts with between 10 and 1000 transaction records in the range they survey, achieving strong performance of a lightGBM-based model.

Heterogeneous networks have also been used for this purpose. Xu et al.[26] distinguishes between accounts, smart contracts, and other entities in their network. Huang et al.[27] rather than distinguishing between different types of nodes, distinguishes different kinds of transactions. In particular they categorize external ET, internal ETH, ERC-20, ERC-721, and ERC-1155 transactions and are able to effectively identify phishing accounts.

Fan et al. [9], which focuses on smart contract scams, builds the "Account Interaction Network", the topological features of which are used to train a classifier in addition to contract bytecode.

These myriad scam detection approaches have proven effective, however ERC-20 transfers have only been incorporated into some of these models, and focus remains on how they are used by accounts rather than their characteristics in and of themselves as we investigate here.

## 3. DATASET AND METHODOLOGY

### 3.1. Data

The transfer events required by the ERC-20 standard represent exchanges of tokens between users. When a token from contract $C$ is sent from user $A$ to user $B$, user $A$ calls the transfer function of $C$. Upon success the balances of both users are updated, and an event is emitted. Whenever a transfer occurs, the ERC-20 standard requires contracts to trigger a Transfer event that contains the addresses of users $A$ and $B$ and the amount transferred. These events are logged in the public chain, and the signature that identifies them is shown here:

```
event Transfer(address indexed _from, address indexed _to, uint256 _value)
```

These transfers may be triggered by the user, as in the above example, or by an entity that user $A$ trusts, such as a smart contract implementing a decentralized exchange. In this way the transfers form the basis of the Ethereum token ecosystem.

To obtain transfers, we call an Ethereum node's JSON-RPC API to fetch all logs that match the transfer event signature and have the correct number of fields. We have retrieved all token transfers between blocks 18M (August 26, 2023) and 20M (June 01, 2024) and divided them into spans of 100K blocks. Each span corresponds to a period of just under two weeks.

Using the transfer records for each token, we build graphs, considering the wallet addresses as nodes and the transfers as edges between them. Each token corresponds to one graph for each period in which it is actively transferred.

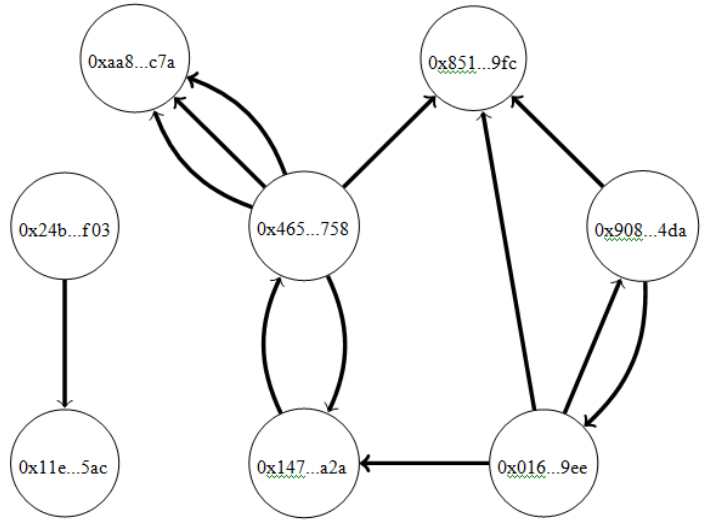

Figure 1. Example token transfer graph

### 3.2. Graph Construction

As we are analyzing properties of token graphs, it is useful to define what is meant by a token transfer graph. The token transfer graph of contract $C$ is a directed multigraph constructed from the Transfer events emitted by $C$, where each transfer is represented by a directed edge between nodes representing its `from` and `to` fields. The formal definition we use can be found in more detail in [21]. Figure 1 shows an example of the structure of such a graph. Two nodes may have multiple, one, or none edges between them depending on whether or not they have exchanged a particular token.

### 3.3. Labeling

In order to define the ground truth for the behavior of token contracts, we collected labels to indicate tokens that are connected with potentially suspicious behavior. Considering web scraping limitations, we limit this collection to token transfer graphs with over 500 nodes in order to focus on tokens with a higher number of users. The sites etherscan.io and honeypot.is were used for this purpose. As a result of this task, we added to the dataset binary "Suspicious" labels for all tokens for which more than 500 distinct wallets have interacted. The totals for this collection are shared in table 1.

Of the 8,389 unique contracts classified, 3,634 or 43% of them were identified as suspicious on at least one of the platforms. When taking the net sum across all periods, however, only 23.9% are suspicious. The reason for this discrepancy is that legitimate contracts are present in more than one period, whereas scams have overwhelmingly shorter lifetimes, so when these duplicates are removed, the proportion of suspicious contracts increases.

Table 1. Overview of transfer dataset.

| Block Range | Accounts | Transfers | >500 nodes | Suspicious |
|---|---|---|---|---|
| 18.0M-18.1M | 2,761,438 | 15,041,939 | 926 | 327 |
| 18.1M-18.2M | 2,553,813 | 13,936,687 | 804 | 257 |
| 18.2M-18.3M | 2,554,730 | 13,690,475 | 775 | 233 |

| 18.3M-18.4M | 2,683,838 | 14,151,825 | 756 | 223 |
| 18.4M-18.5M | 2,497,527 | 15,232,538 | 910 | 239 |
| 18.5M-18.6 M | 2,529,151 | 15,579,423 | 911 | 178 |
| 18.6M-18.7M | 2,667,773 | 15,089,141 | 988 | 214 |
| 18.7M-18.8M | 2,600,629 | 14,947,717 | 984 | 216 |
| 18.8M-18.9M | 3,416,925 | 14,954,427 | 935 | 243 |
| 18.9M-19.0M | 3,282,510 | 14,700,332 | 895 | 224 |
| 19.0M-19.1M | 3,999,995 | 15,603,459 | 884 | 221 |
| 19.1M-19.2M | 2,918,097 | 13,845,342 | 828 | 215 |
| 19.2M-19.3M | 2,812,750 | 14,082,075 | 955 | 228 |
| 19.3M-19.4M | 3,022,560 | 15,830,761 | 1,049 | 218 |
| 19.4M-19.5M | 3,946,732 | 18,145,738 | 1,251 | 316 |
| 19.5M-19.6M | 3,664,400 | 18,230,533 | 1,258 | 295 |
| 19.6M-19.7M | 3,469,843 | 19,120,308 | 1,217 | 288 |
| 19.7M-19.8M | 3,118,480 | 16,992,067 | 980 | 193 |
| 19.8M-19.9M | 3,131,170 | 16,744,683 | 882 | 136 |
| 19.9M-20.0M | 2,662,568 | 18,165,738 | 917 | 101 |

### 3.4. Features

To analyze the token transfer graphs, we compute several features for each network. **num_nodes** and **num_edges** denote the number of nodes and edges respectively in the graph. **density** is the ratio between the number of edges in the graph and the maximum possible number of edges. Note that this value may be greater than one, since token transfer graphs are directed multigraphs. **num_components** is the number of weakly connected components in the token transfer graph. Two nodes are weakly connected if there is some path connecting them, ignoring the direction of edges. A weakly connected component then is a maximal subgraph of the token transfer graph consisting of connected nodes. **avg_comp_size** is the mean number of nodes that compose each weakly connected component. **lifetime** is the time between the first and last observed transfer. **transfer_std_dev** is the standard deviation of the distribution of the block numbers of transfer events. **amount** is the total amount transferred in the network, taken as the sum of the amount fields of the transfer events.

### 3.5. Classification Model

With the extracted features, we train a logistic regression model to classify token networks as either tied to scam behavior or not. Logistic regression is a linear model for classification which attempts to fit a sigmoid curve to chosen input variables to predict a categorical output variable (in this case whether the contract was labeled as suspicious). The model finds coefficients for each of the input features, so the final prediction is essentially a weighted combination of these features. Essentially, a linear combination $z$ of the input features $x$ is calculated:

$$z = \beta 0 + \beta 1 x 1 + \beta 2 x 2 + \ldots + \beta n x n$$

A sigmoid mapping function is then applied to convert it into a probability:

$$P(Y = 1 \mid X) = \sigma(z) = \frac{1}{1 + e^{-z}}$$

Finally, a threshold of (0.5) is applied to classify the observation into one of two classes to generate a prediction:

$$Y'(P) = \begin{cases} 1 \text{ if } P \geq 0.5 \\ 0 \text{ otherwise} \end{cases}$$

The model is trained and evaluated based on metrics of accuracy, precision, recall, and F1-score, which can be calculated given the number of true and false positives and negatives classified by the model.

Feature extraction and model training and evaluation were performed on a standard laptop with 16 GB of RAM. The transformation of JSON-RPC output and extraction of features was accomplished in approximately two hours for all 314M transfers. The training and evaluation of the simple regression model was achieved in less than a quarter second on average for each of the 20 block ranges.

## 4. RESULTS

### 4.1. Feature Distribution Patterns

The feature distributions in figure 2 offer certain insights into the differences between scam and legitimate contracts. The distributions of both `lifetime` and `transfer_std_dev` have a majority of scam contracts with lower values while valid contracts last longer and are less clustered near the mean. This validates intuition about temporal behavior of illegitimate contracts; low standard deviation for transfer events means that transactions are more clustered, rather than evenly spread out over the activity period as would be expected for a token being regularly used in decentralized exchanges and commerce, and short lifetime indicates that scams will operate for periods less than 10,000 blocks (or about a day and a half), before being discovered, ceasing to be profitable, or moving on to new targets. Scam networks are also less dense than legitimate ones, because though the scam token may be distributed to many users, many will not engage with the scams.

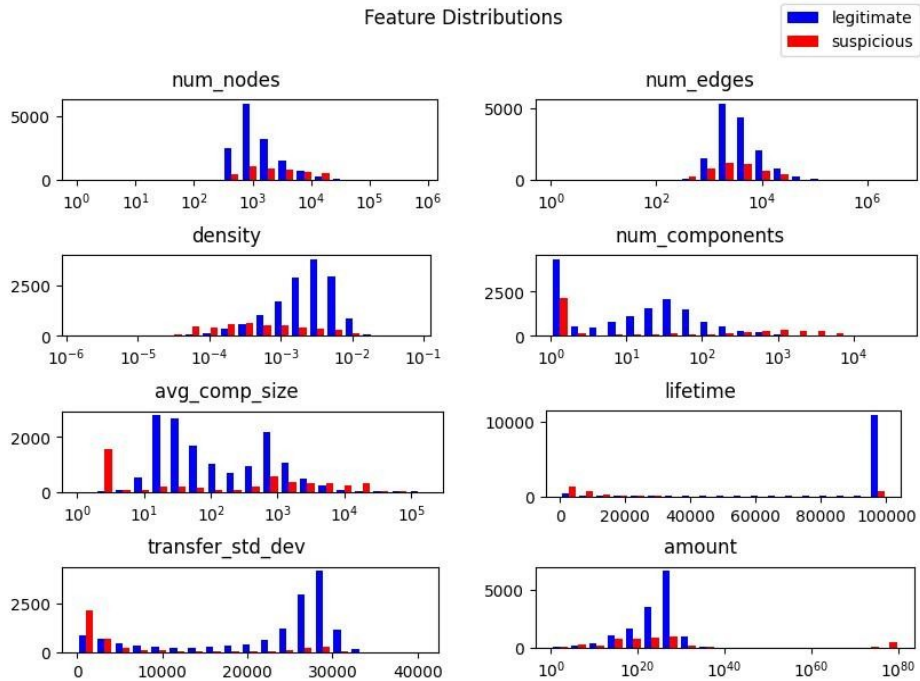

Figure 2. Feature distributions

Additionally, scam networks are more likely to have either very large or very small components, which could indicate the existence of different scam mechanisms. The networks with average components of two or three nodes are potentially so-called "address poisoning scams" [28], where scammers will send tokens from addresses similar to ones the victim interacts with often, resulting in scattered graphs of small components because each victim will interact with different people. The scams with one large single component, meanwhile, seem to be centered around decentralized exchanges, implying that the scam is a honeypot or rug pull, seeking to entice investors before removing liquidity.

## 4.2. Classification of Particular Token Transfer Networks

To further substantiate observations about the feature profiles of scam and legitimate contracts, we present a comparison of similarly sized graphs falling into the various patterns of behavior observed.

Figure 3 shows the token transfer graphs for four smart contracts from the range 18.0M-18.1M. 3a, 3c, and 3d all have close to 2000 nodes, and illustrate the three observed network structures for these token transfer networks.

3a, depicts the BAT token used by the Brave web browser and holds the structure shared by most of the legitimate networks observed, with a large and interconnected main component of 1694 nodes (over 75% of the network), surrounded by smaller components of only a few nodes. In contrast, 3c consists of a single large star-shaped component. In the network of 2,173 addresses, only two have degree greater than 3, those being the null address, associated with events that burn and mint the token, and the Uniswap liquidity provider token, that allows Shiacoin to be exchanged on decentralized exchanges. This indicates that nearly all activity of this token is taking place on a single decentralized exchange, most likely in some form of rug pull or honeypot scam. The structure of 3d, a copycat USDC token, is characteristic of counterfeit scams, with lots of small components averaging no more than three or four nodes each. Scams such as these send copycat tokens to users from addresses close to those familiar to them, which explains why scammer would send the token from a multitude of addresses, rather than a single one.

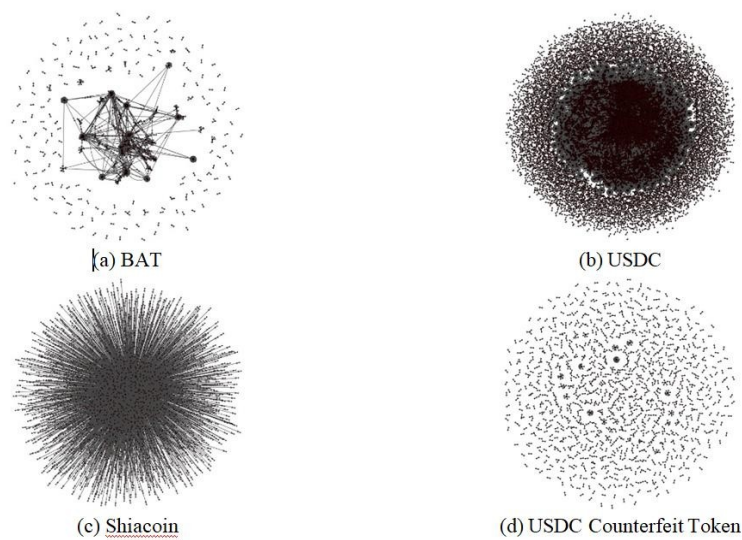

Figure 3. Examples of token transfer networks

3b is the network for USDC, the legitimate counterpart to the counterfeit token, and in addition to being much larger than the counterfeit token network, it also has a different structure close to that of 3a and the other legitimate networks, showing that the patterns seen in the smaller networks can scale to the networks of larger tokens.

## 4.3. Classification Model

To evaluate the applicability of these qualitative observations, we use these features to train a logistic regression model to classify the networks in each block range.

We used 5-fold cross-validation to evaluate the model while accounting for potential overfitting bias, taking the average values for accuracy, precision, recall, and F1-score across the five runs. The average accuracy for all 20 block periods was 88.7%. The average precision and recall were 86.4% and 62.0%, with the average F1-score at 71.2%. Particularly in the F1 measure, the model declines in more recent block periods, but maintains high accuracy, predominantly due to a steep drop off of recall, as seen in figure 5.

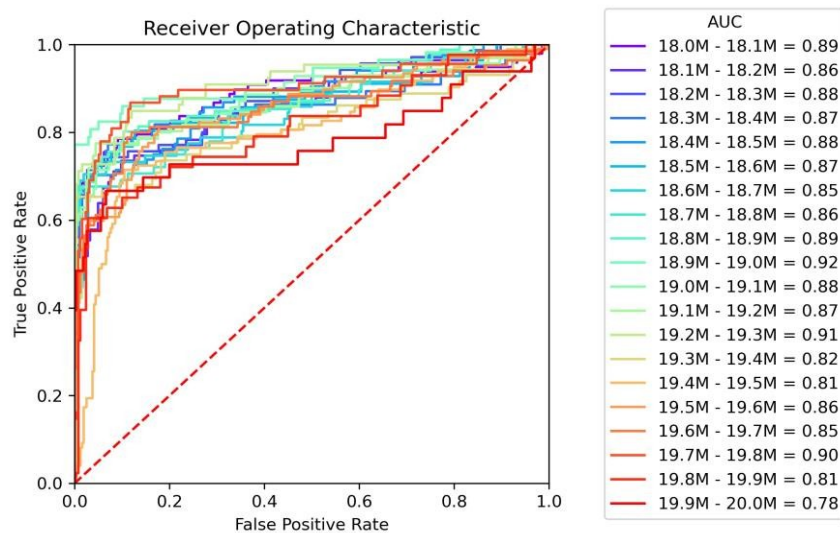

Figure 4. ROC curves for each block range

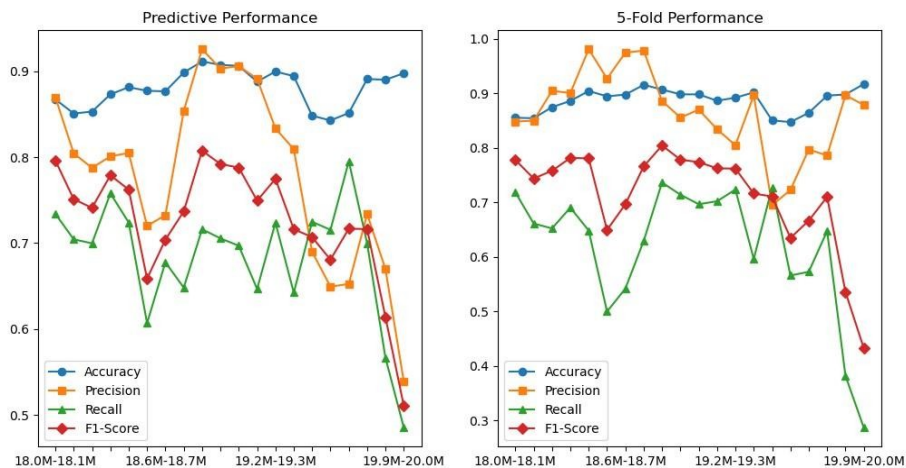

Figure 5. Model performance in each block range

The model averaged an AUC of 0.86 for the 20 block ranges. The ROC curves for each block range are shown in figure 4 and have no significant differences between themselves.

The model was also able to make predictions for different time windows. A logistic regression model was trained using the 926 labeled contracts in the first block range, and then evaluated on each of the other 19 block ranges using the same metrics as in the 5-fold cross-validation.

The average accuracy and F1-scores of 88.0% and 72.5% respectively are extremely similar to those of the model when evaluated with data from the same block range, and both accuracy and F1-score follow similar trends when compared with those of the 5-fold cross-validation, as seen in figure 5. This performance shows promise that these network characteristics retain their classification power and that this behavior is not unique to short or specific periods.

The difference in the behavior of precision seen in figure 5, can be attributed to the class imbalance for later blocks compared to the initial periods. The model trained on blocks 18.0M - 18.1M, where a third of all labeled contracts were scams has many false positives in the recent spans where only one ninth of labeled contracts are suspicious. Meanwhile the model trained in the period 19.9M - 20.0M has fewer problems with generating as many false positives.

### 4.4. Analysis of False Positives and Negatives

We have selected some false positives and negatives from one of the runs of 5-fold cross validation in the range from block 18.0M to 18.1M. Though not comprehensive, understanding some of the characteristics that cause the model to falter is a step to improving it in future work.

#### 4.4.1. False Positives

> **0x3983c181b68d53b2dd9024798028bddac45b192d**: This address is associated with the token KRESTK. It was likely flagged by our model due to consisting of a single component and having a relatively short lifetime of 24,909 blocks.
> **0xbf94a2ceeaa256a0444970176813be7841147f18**: This address is associated with the token REDHC. It shares similar parameters to the above address, with a single component and a lifetime of 6507 leading to its classification as a scam. A difference between the two tokens revealed by further inspection is that this one maintains liquidity, while the other does not.

The false positives, including those not discussed here, all share similar characteristics, among them short lifespans and consisting of a single component, which likely led to their misclassification. It is also difficult to establish whether or not they are scams; though they did not meet the threshold used for the construction of our dataset, it is possible that they engage in malicious behavior and should be classified as such.

#### 4.4.2. False Negatives

> **0x7721a4cb6190edb11d47f51c20968436eccdafb8**: This token, GUISE, was labeled as a scam in the ground truth dataset due to its being flagged by honeypot.is for a high sell tax and low liquidity. Though there are traces of a project associated with it, the website is down, and it shows signs of being a rug pull scam. Its lifetime of 99,952 blocks likely contributed to its classification as legitimate, as well as the fact that its graph consists of multiple components.

>    **0x064bac371767c7ef2bdaaf8755eb603e1aa9189e**: This address holds a contract hosting a counterfeit token for Tether USD, one of the most popular tokens on the Ethereum network. Unlike most of the scam contracts, it had a long lifespan of 99,932 blocks.
>    **0xc46c407570eff0e9a5f6afcf623039436c1e4497**: This token, 7-Eleven (711), was classified as legitimate, perhaps in spite of its short lifespan of 23,598 blocks. There is little information about the token, though its low liquidity is what led to its classification as a scam. This is a profile that matches many of the false negatives, and also illustrates the difficulty of establishing a ground truth dataset when many tokens have little to no information about them circulating.

Analysis of these false classifications reveals that the model may place over importance on lifespan leading to some of these issues, as well as illuminating some of the challenges with constructing a true ground truth dataset. In future work we may improve this analysis by incorporating more complex features and methodologies that could improve the accuracy and reliability of the model. By integrating advanced machine learning techniques or feature engineering approaches, such as examining deeper tokenomics and leveraging network graph analytics, we might be able to better capture the nuanced characteristics of these tokens.

### 4.5. Predictions for Unclassified Networks

Though we lack a ground truth for token transfer networks with fewer than 500 nodes, our model can make predictions about the legitimacy of those smaller tokens. To avoid extrapolating beyond training data with size related parameters, we remove `num_nodes`, `num_edges`, and `num_components` from the model and add a new feature, `edges_per_component`, to describe the average number of edges per component in each graph and compensate information lost by removing the three other features.

The logistic regression model is re-trained on all 19,105 labeled token transfer networks, of which 23.9% are labeled as scams. The model with the adjusted features performs similarly to the original model, with an average of 86.2% accuracy and an F1-Score of 69.0% across the 5-fold cross validation. When applied to the 613,249 token transfer networks with under 500 nodes, the model predicts that 372,931 of them are scams, or 60.8%, far above the 23.9% in the training set. Further investigation reveals that this is due to the outsized effect of small networks with short lifetimes. Networks with more than 100 nodes (of which there are 38,042) have scams predicted at a much lower rate of 25.4%. Networks with lifetimes under 1000 blocks have scams predicted at a staggering rate of 99.7%.

When lifetime is removed from the model as well, performance takes a small hit down to 85.6% accuracy, while the number of predicted scams plummets to 11.7%. While this emphasizes the important role of contract lifetime as a predictive feature, it also highlights it is not the only discriminant factor in our model.

These numbers paint two contrasting perspectives of Ethereum token transfer networks. First, there is the view that the majority of tokens are legitimate, with scams predominantly found among larger networks. These larger networks are either successful scams or are attempting to deceive a significant number of users. Alternatively, there is the perspective that most smaller networks are scams, while legitimate tokens naturally expand and reach a larger audience, growing in size and scope. Without annotated ground truth data for these smaller tokens, it is challenging to determine which perspective is more accurate, as both scenarios account well for the observed model accuracy in larger token transfer networks.

The lack of a ground truth dataset for smaller networks also leads to challenges for catching scam behavior in early stages of the network, rather than after the fact. Younger networks will be smaller by nature and understanding more about how scams look at smaller scales would be helpful for developing real-time scam detection. Exploring this issue further will be left for future work, where the focus will be on establishing a comprehensive ground truth to better understand the nature of smaller token networks.

### 4.6. Discussion

The model presented here, particularly when considering its relative simplicity, is seen to perform comparably to others in the literature. For example, though the approach with an LGBM model achieves impressive performance, a logistic regression model presented with the same features achieves accuracy of 86.69% [4], comparable to our logistic regression model.

Models that outperform ours tend to employ more sophisticated architectures, such as an XGBoost model that attains 96.82% accuracy by leveraging 10 features tailored for predicting smart contract scams, which aligns closely with our use case [6]. Other XGBoost models report 96.3% accuracy with 42 features [5], and between 88.9% and 94.6% accuracy across various datasets using 8 distinct graph-topological features [9].

These comparisons underscore that while more advanced models and a greater number of extracted features can lead to enhanced performance, our model's simplicity is, in fact, a significant advantage. It effectively identifies new patterns in the behavior of scam tokens without the need for complex methods. This results in an outstanding cost/benefit ratio, as it yields strong predictive capabilities at a fraction of the complexity and computation cost of other sophisticated methods, making it a practical and efficient solution in the context of detecting fraudulent activities.

## 5. CONCLUSIONS

Our research demonstrates that token transfer networks possess structural characteristics capable of predicting network behavior, with these traits evident across multiple time intervals. Through the analysis of 20 segments of 100,000 blocks each, we identified patterns in suspicious tokens that effectively predict the nature of these networks. Our method shows notable efficacy in detecting suspicious tokens within Ethereum networks, achieving an accuracy rate of 88.7% during evaluation. This high level of performance underscores the strength of our approach, particularly in balancing simplicity and effectiveness. By leveraging key structural features of token transfer networks, our method excels in distinguishing between legitimate and fraudulent activities without the need for complex models.

This study is not without limitations, as it relies on scam reports from platforms like etherscan.io and honeypot.is. Etherscan, in particular, depends heavily on community reporting, while Honeypot has acknowledged its own limitations. Consequently, our dataset may lack completeness, especially concerning newer contracts. Nonetheless, our approach boasts an impressive cost/benefit ratio. By employing a relatively simple logistic regression model, we achieve strong predictive performance without resorting to complex architectures or extensive computational resources, making it both accessible and cost-effective compared to more sophisticated models.

To further advance this research, we have introduced a new dataset that provides a robust foundation for analyzing token transfer networks. Moreover, we are releasing the code from our experiments to foster reproducibility and encourage further exploration by the research community. Future research will focus on improving data collection and validation methods, such as collaborating with blockchain security firms for better ground truth data and exploring machine learning techniques to identify scams based on on-chain behavior. Additionally, examining token transfer networks over shorter timespans could facilitate quicker scam detection, with sliding window approaches potentially offering real-time insights that can be applied in up-to-the-minute scam detection.

Moreover, incorporating advanced network analysis techniques can significantly improve predictive accuracy. Future studies will explore temporal network metrics to capture dynamic behavior within token networks, and implementing community detection algorithms could help identify suspicious subgroups. While our logistic regression model proved effective, investigating more advanced machine-learning techniques, such as graph neural networks, might further enhance prediction capabilities.

Expanding research to include cross-chain analysis presents another valuable opportunity. By comparing token network structures across different blockchain ecosystems—like Ethereum, Binance Smart Chain, and Solana—we can identify common scam patterns and develop unified detection methods applicable across platforms. These efforts will build upon the foundation laid by this study, ultimately contributing to the security and trustworthiness of decentralized finance and token economies.